\documentclass[11pt]{article}
\usepackage{epsf}
\usepackage{graphicx}        
\def\c{\cite}

\def\r1{(\ref{$1})}

\def\ll{\label}

\def\ba{\begin{array}{c}}

\def\ea{\end{array}}

\def\l{\left}
\def\l({\left(}
\def\r){\right)}
\def\r{\right}

\def\be{\begin{equation}}
\def\bc{\begin{center}}
\def\ec{\end{center}}
\def\bit{\begin{itemize}}
\def\eit{\end{itemize}}
\def\ee{\end{equation}}
\def\ed{\end{document}}
\def\bea{\begin{eqnarray}}
\def\eea{\end{eqnarray}}
\def\efr{\end{flushright}}

\begin{document}
\title{{Integrable inhomogeneous NLS equations are equivalent to the standard 
NLS  }}
\author{ Anjan Kundu\\ 
 Theory Group, 
Saha Institute of Nuclear Physics\\
 Calcutta, INDIA\\[10mm]
{anjan.kundu@saha.ac.in}}
\maketitle
\begin{abstract}
A class of inhomogeneous nonlinear Schr\"odinger equations (NLS), 
claiming to be novel integrable systems with  rich properties 
continues appearing in PhysRev and PRL.
All such equations are shown to be not new but equivalent to the standard
NLS, which trivially explains their integrability features.  
 \end{abstract}


Time and again  various forms of inhomogeneous nonlinear Schr\"odinger
 equations (IHNLS) 
along with their discrete variants are appearing  as  central result
 mostly in the pages of Phys. Rev, and  PRL 
 \c{CLU76,RB87,PRA91,Kon93,PRL00,PRL05,PRL07}, which are either suspected to be
 integrable due to the finding of  particular analytic   or stable computer
 solutions, or assumed to be only Painlev\'e integrable \c{arxiv08}, or
 else 
claimed to be completely new integrable systems.
Apparently  the solution of such integrable systems needs  
generalization of the   inverse scattering method (ISM), in which 
 the usual { isospectral}   approach 
 involving only   constant spectral parameter
$ \lambda$ has to be extended to 
  {\it nonisospectral} flow with time-dependent
$ \lambda (t)$.  Moreover  certain features of the soliton solutions 
of such inhomogeneous NLS, like the changing of the  solitonic   amplitude, shape  and
velocity  with time were thought to be  { new and surprising 
  discovery}.

We show here that all these  IHNLS 
, though completely integrable are not  new or independent
integrable systems, and   in fact are equivalent to the standard 
 homogeneous NLS,
linked through simple gauge, scaling and coordinate transformations.
The standard NLS   is a  well known 
 integrable system with known Lax pair, soliton solutions and usual
isospectral ISM \c{nls,ALM}. As we   see below,  a simple 
time-dependent gauge 
transformation  of the standard isospectral
 system with constant 
$\lambda $ can  create the illusion of having complicated 
nonisospectrality. Similarly, a time-dependent scaling of the standard 
NLS field $Q \to q =\rho(t) Q   $  would naturally   lead the 
  constant  soliton amplitude  to a time-dependent one. In the same way
a trivial coordinate transformation $x \to X=\rho(t) x   $
 would change the usual constant 
 velocity $v$ of the   NLS soliton  to a time-variable quantity $v(t)=\frac
 {v} {\rho(t)} $  
 and 
the invariant
shape  of the standard soliton with constant extension $\Gamma=\frac 1 {\kappa}$ to a
time-dependent  one
with variable extension $  \Gamma (t)= \frac \Gamma {\rho(t)}
$ (see Fig 1a a,b).
Therefore all the rich integrability 
 properties of the IHNLS, observed   in earlier papers,   
 including more exotic and seemingly   surprising 
features like
nonisospectral flow,
 appearance of shape changing and accelerating soliton etc. can be
trivially explained from the  time-dependent transformations of these IHNLS
from the standard NLS and 
   the corresponding 
 explicit  result , namely     the  Lax pair, N-soliton
solutions, infinite conserved quantities  etc. for the inhomogeneous NLS models
  can be  derived easily
   from their well known counterparts in the homogeneous  NLS case through
the same transformations \c{nls}.

Let's start from a recent version of IHNLS \c{PRL07}, which is generic in
some sense:
\be iQ_{t}+\frac 1 2 D Q_{xx}+R|Q |^2Q -(2\alpha x+ {\Omega^2}  x^2)Q=0
 , \ll{ihnls1}\ee  
where the nonautonomous coefficients $D(t),R(t) $ of the dispersive and the nonlinear
terms  are arbitrary functions of $t$ and the other time-dependent 
functions are \be
\alpha (t)=s_t\rho , \ \  \Omega^2(t)= \frac 1 4 (
\theta_t-\frac 1 2 D\theta^2), \  \mbox{where}
 \  \rho=\frac R D, \   \ \theta=\frac {\rho_t}{\rho D} 
 ,\ll{coefft} \ee 
  $s(t) $ being another
arbitrary function. It is easy to see that a  time-dependent  scaling 
  of the field can change the coefficient
 of the nonlinear term in (\ref {ihnls1}) and at the
same time generate an additional term from $iQ_t $, while 
a change in phase of the field involving $x^2$
 would yield extra terms from $Q_{xx} $.
As a result transforming $Q \to q=\sqrt \rho e^{i \frac \theta 4 x^2 }Q $.
we can rewrite IHNLS (\ref{ihnls1}) into another form
\be iq_{t}+\frac D 2 q_{xx}+D|q |^2q -(2\alpha x+iD \theta)q-iD \theta x
q_x=0.
  \ll{ihnls2}\ee  
In \c{PRL07}
Eq (\ref{ihnls1}) was declared to be a new discovery and as a proof of its
integrability a Lax pair associated with Eq (\ref{ihnls2}) was presented,
which we rewrite here in a compact and  convenient form by introducing a matrix 
$U^{(0)}(q)= \sqrt \sigma (q \sigma^+-q^* \sigma ^-) $ as
\bea
U(\lambda(t))&=&-i\lambda(t) \sigma^3+U^{(0)}(q),  \nonumber \\
 V(\lambda(t))&=& D V_0(\lambda(t))-i  \alpha  x \sigma^3+D \theta x
U(\lambda(t)))
,\ll{UVt} \eea
where  
\be V_0(\lambda(t))=-i\lambda^2 (t) \sigma_3 + \lambda (t) U^{(0)} + \frac i
2 \sigma_3 (U^{(0)}_x -
(U^{(0)})^{2}). \ll{Vt} \ee 
We can check from the above Lax pair  that 
the flatness condition 
$U_t- V_x+[U,V]=0$  yields the IHNLS (\ref{ihnls2}) under the  
 constraint $ \lambda(t)_t=\alpha+ D \theta \lambda (t)$. Using relations
(\ref{coefft}) one can resolve this constraint to get   $\lambda (t)=\rho
(t)(\lambda + s(t)) $, which was given   in \c{PRL07}.
We  now establish the equivalence between the  Lax pair 
$U(\lambda(t)), V(\lambda(t))$  (\ref{UVt}, \ref{Vt})
for the IHNLS and  the well known Lax  pair 
$U_{nls}(\lambda), V_{nls}(\lambda) $ of the  standard  NLS  \c{nls},
  showing  explicitly that the 
nonisospectral $\lambda(t) $ is  convertible to constant spectral parameter 
$\lambda $ through simple transformations. 
For this 
	it is interesting to  notice first, that the  structure 
of the NLS Lax pair  is   hidden already  in the  expression of the 
IHNLS  Lax pair
as 
$U(\lambda(t)=\lambda)= U_{nls}(\lambda)$ and $ V_0(\lambda(t)=\lambda)=
V_{nls}(\lambda)$.
 Therefore the  aim should be  to remove the $t$-dependence from $\lambda (t)=\rho
(t)(\lambda + s(t)) $ by absorbing the arbitrary functions
 $ \rho(t) $ and
$s(t)$ in  step by step manner.
Note   that the Lax pair $U,V $,
 as evident from the associated linear
problem $ \frac \partial {\partial x}\Phi=U\Phi, \ 
\frac \partial {\partial t} \Phi=V\Phi $, correspond  to 
 infinitesimal generators in the
$x$ and the $t$ direction, respectively and therefore    a simple
 coordinate change $ (x, t) \to (\tilde x=\rho(t)x, \ \tilde t = t)  $ resulting $
\frac \partial {\partial \tilde x}=\frac 1 \rho \frac \partial {\partial
x}, \ \frac \partial {\partial  t}=\frac \partial {\partial \tilde t}+
 {\rho_t}  x\frac \partial {\partial \tilde
x}=\frac \partial {\partial \tilde t}+
 D \theta   x\frac \partial {\partial 
x} $, would yield $U( x, t)=  \rho  U(\tilde x,\tilde t) $
and $V( x, t)= V(\tilde x,\tilde t)+ D \theta x  U( x, t) $. Therefore
using such a transformation and comparing with  (\ref{UVt}), 
we can easily remove the $\rho(t)$ factor from $
\lambda (t)$ in $U(x,t) $, which however would scale the field as $q \to
\frac q \rho $ and   at the same time eliminate from the transformed $V(\tilde x,\tilde t) $ the  
 nonstandard term  $ D \theta x  U( x, t) $ appearing in V( x, t) (\ref{UVt}).
 For the
removal of additive term $\rho (t) s(t) $ from $\lambda (t) $, present  
in  $U( x, t) $, one can perform a gauge transformation $\Phi \to
\tilde \Phi = g\Phi $ with $ g=e^{i \rho s \sigma^3}$,
taking the Lax pair to a gauge equivalent pair
\be \tilde U=g_xg^{-1}+g Ug^{-1}, \ \tilde V=g_tg^{-1}+g Vg^{-1}. \ll{GT}\ee
 One notices  that though the above transformations are enough to
remove explicit $ t$ dependence from $U $ due to its  linear dependence 
on     
 $\lambda (t)$, the removal of $ t$ from   $V(\lambda(t)) $ 
becomes a bit  involved due to the nonlinear entry of
 $\lambda ^2 (t)$ 
and $ \lambda (t) U^{(0)} $ in it, which
bring in more time-dependent terms  like $2D\rho ^2 s $ and $ D\rho ^2 s^2 $. 
 These extra terms however can be exactly compensated for 
by  extending slightly  the above coordinate and gauge transformations  by 
 introducing additional functions $f(t), \ \tilde f(t) $
and choosing them as   $f_t=2D\rho ^2 s $ and $  \tilde f_t=D\rho ^2 s ^2 $.
The multiplicative factor $D\rho ^2 $ appearing in all 
terms in  $V(\lambda(t)) $  can be  absorbed easily  by a further coordinate change 
$t \to T=D\rho^2 t $.
Therefore taking the above arguments into account
  one finally  solves the problem completely through the following
three steps of simple transformations:\be \noindent
i) \mbox {{\it  Coordinate transformation}}: (x,t) \to (X,T), 
 \ \  X=\rho(t)x +f(t), \  T=D\rho ^2 t, \
\mbox{with} \ f_t=2D\rho ^2 s,
\ll{XT}  \ee
\be ii) \mbox{{\it Gauge transformation}} \  (\ref{GT}),   
\  \mbox {where} \   \ g=e^{i (\rho(t) s(t)x+\tilde f(t)) \sigma^3}
\ \mbox{with} \ \tilde f_t=D\rho ^2 s^2,
\ll{g}  \ee
\be iii) \mbox{{\it Field  transformation}}:  q \to \psi, \
 \mbox{where} \  \psi= \frac 1 {\rho (t)}
   q \ e^{2i(\rho (t) s (t) x+\tilde f(t)) }.
\ll{psi}  \ee
The above transformations would  take 
  (\ref{UVt}) directly to the standard NLS Lax pair
\bea
U_{nls}(\lambda)&=&-i\lambda \sigma^3+U^{(0)},\
 \mbox {where} \ U^{(0)}=\psi \sigma^+-\psi ^* \sigma^- ,  \nonumber \\
  V_{nls}(\lambda)&=&
-i\lambda^2  \sigma_3 + \lambda  U^{(0)} + \frac i
2 \sigma_3 (U^{(0)}_X -
(U^{(0)})^{2}). \ll{UV} \eea 
 which  proves  the equivalence of 
the Lax pair (\ref{UVt}) for the IHNLS (\ref{ihnls2})  and  the Lax pair (\ref{UV})
associated with 
 the standard NLS:
\be i\psi_{T}+\frac 1 2 \psi_{XX}+|\psi |^2\psi 
=0, \ll{nls}\ee
obtained as  the flatness condition of (\ref{UV}).
One can also  check  that under the change of independent and dependent
variables (\ref{XT}) and (\ref{psi})  the inhomogeneous NLS (\ref{ihnls2}) is
transformed  
directly  to the homogeneous  NLS (\ref{nls}).

Therefore we remark that   the inhomogeneous NLS (\ref{ihnls1})  and
 (\ref{ihnls2}) are equivalent to the homogeneous NLS
(\ref{nls}),  a  well known  integrable  system. The corresponding  Lax pairs (\ref{UVt}) and (\ref{UV})
are also gauge equivalent to each other, which therefore
 trivially explains the  complete   integrability of the inhomogeneous NLS.    
  All signatures of  the complete  integrability    like the  Lax pair, N-soliton
solutions, infinite conserved quantities  etc.  for these HNLS can be  obtained
easily from the
corresponding  well known  expressions for the NLS system (\ref{nls}, \ref{UV}) by
inverting the set of transformations (\ref{XT},\ref{g},\ref{psi}) as  $ (X,T)
\to (x,t) , \ g \to g^{-1}, \ \psi \to q $.
As a result,  explicit $ t$-dependence 
obviously enters in the  Lax
operators as well as in the amplitude, phase and the x-dependence of the
field $ q(x)$ of
the IHNLS system,
  resulting the spectral parameter $\lambda \to \lambda (t) $ and  making  the
 constant  amplitude $A$, extension $\Gamma $ and  velocity $ V$  of the
 soliton to become
$t$-dependent.  
Fig. 1  demonstrates this situation, showing that the NLS soliton (module) a) 
 $|\psi |=A {\rm sech} \xi, \ \xi=\frac {1} \Gamma (X-VT) $ goes to IHNLS
soliton  b):  $|q |=A(t)  {\rm sech} \tilde \xi, \ \tilde \xi = \frac {1}
{\Gamma(t)} (x-v(t))$, where  $ A(t)=A \rho (t)$,  $v(t)=D V \rho(t) t-\frac { f(t)} {\rho (t)}, 
\Gamma(t)=\frac {\Gamma} {\rho (t)}$ under the transformations  inverse to
(\ref{XT},\ref{g},\ref{psi}).
Therefore even though IHNLS soliton (Fig. 1b)  looks rather exotic and 
quite different from the standard NLS soliton (Fig. 1a), these solutions 
 are  related simply 
by coordinate and scale transformations and 
 belong to
equivalent integrable systems.

\begin{figure}[c]

\includegraphics[width=5.cm,height=4.1 cm]{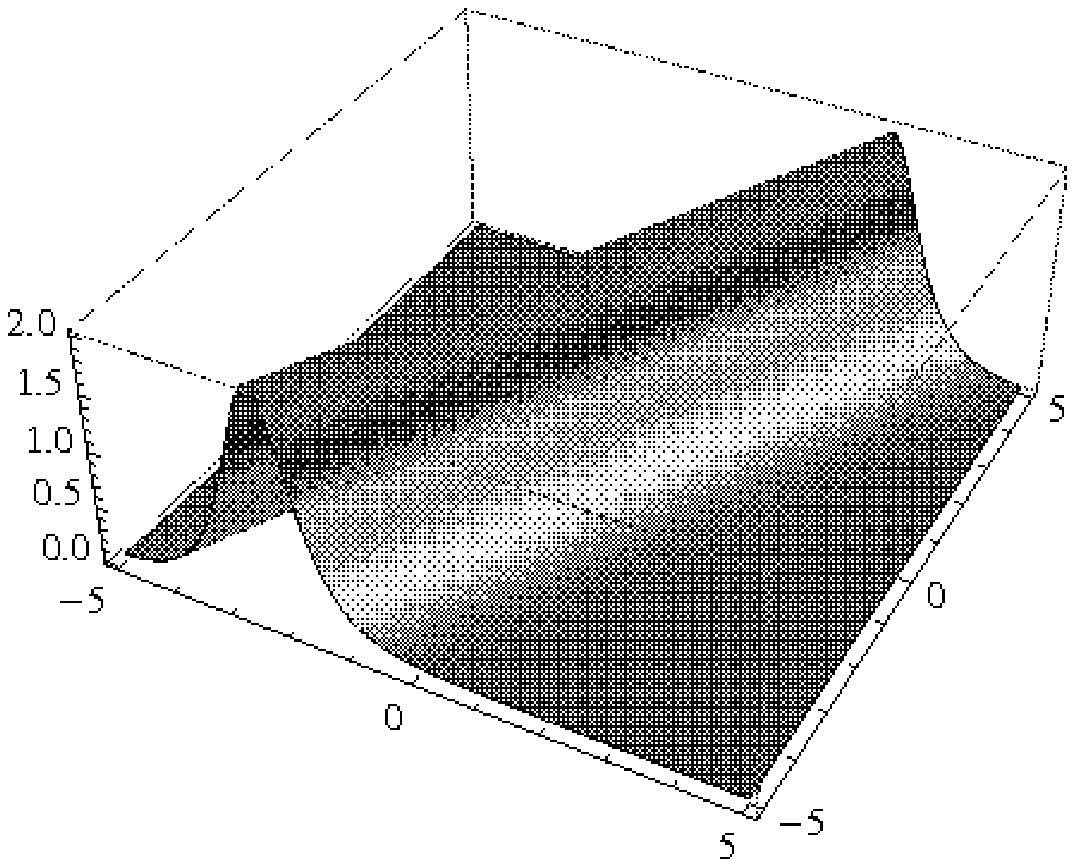}
 \quad \quad \quad \quad \quad \quad \quad \quad 
\includegraphics[width=5.cm,height=4.1 cm]{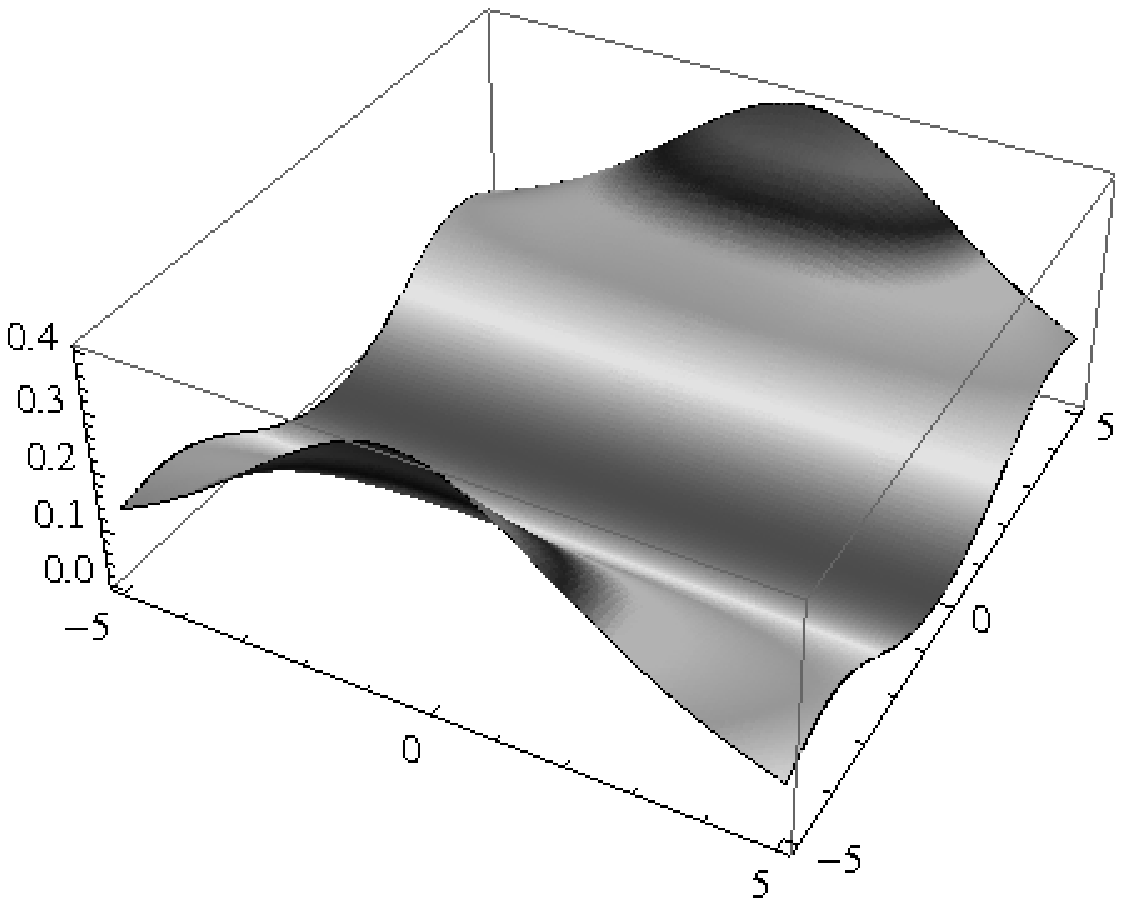}
 
\caption{  Exact soliton
 solutions (module)
  for  the integrable a) homogeneous NLS  (\ref{nls}) and b) inhomogeneous NLS (\ref{ihnls2}) with
 $A=2.0, V=0.5 $. and  the particular choice $R=.008t^2, D=1, s=0 $.
In spite of the significant differences  between the appearance and dynamics of
these two solutions they are related by simple transformations
(\ref{XT}-\ref{psi}) and belong to 
equivalent integrable systems.
}
\end{figure}

It is worth mentioning that, though in all earlier papers only 1-soliton of
the IHNLS was considered,  one can easily derive  the exact
N-soliton for the IHNLS, thanks to its complete integrability, by exploiting again
 its equivalence with the  integrable NLS, i.e. by simply mapping
the known N-soliton of the standard NLS through the same transformations
(\ref{XT}-\ref{psi}).

By  redefining    the
field further: $ Q\to b (t) Q e^{i a (t)} $ with arbitrary functions $a(t), b(t) $,
 we can generate   more inhomogeneous terms in (\ref{ihnls1}) resulting a more
general form of IHNLS
\be iQ_{t}+\frac D 2 Q_{xx}+R|Q |^2Q -(2\alpha x+ {\Omega^2}
x^2+a+i\gamma )Q=0, \ \ \gamma (t)=\frac {b_t} b,
  \ll{ihnls3}\ee  
 equivalent naturally to  the integrable
NLS.   The   IHNLS (\ref{ihnls3}) was found to be the maximum inhomogeneous NLS
system which can pass the Painlev\'e integrability criteria \c{PLA87}.
A recently proposed   IHNLS \c{arxiv08}, which is simply  a particular case of
(\ref{ihnls3})
at $ a=0 $ and $\alpha =0 $,
 is therefore also equivalent to the standard NLS and 
hence, contrary to the assumption in \c{arxiv08} that the system is only Painlev\'e
integrable  
and {\it not} completely integrable, the equivalence with 
 NLS assures  the complete integrability,
including the existence of infinite conserved quantities, N-soliton solutions etc.
  for this IHNLS \c{arxiv08}

 We now look into other forms of integrable HNLS appeared earlier  in Phys. Rev. \c{RB87,PRA91,Kon93}
and PRL  
  \c{CLU76,PRL00,PRL05} and show  their  equivalence 
to the standard NLS, similar to as  found above. 
The simplest  form  of inhomogeneity to the NLS: $2x q $ was proposed  in
\c{CLU76}, which is clearly   a particular case of (\ref{ihnls2}) with $ \alpha =1,
\Omega =0$, ensured by the choice   
 $R=D=1, s=t$,  proving thus its equivalence with the NLS.

A more general IHNLS with $F(x) Q $ was considered in \c{RB87} and shown
finally that integrability restricts the choice only upto $F(x)=a+\alpha
x+\mu x^2 $, which is consistent  with the general
integrable IHNLS (\ref{ihnls3}), shown  to be equivalent to the standard NLS
(\ref{nls}).
 However  for constructing such integrable IHNLS, as   shown here, 
 x-dependent 
spectral parameter  considered in \c{RB87} is not needed and  similarly 
the restriction on   function $ h(t)$ appearing in $ \lambda (t)$,
found by the author apparently as a condition  for the integrability, 
  actually does not appear allowing  the function to be arbitrary,  
as shown here.

In \c{PRL00} a variant of IHNLS was considered, which was suspected to be
integrable through computer simulation. It is easy to see however, that this IHNLS  can be
obtained as a particular case 
from  (\ref{ihnls3}) at $\alpha=0, s=1, a=0, \gamma =0, \Omega=0 $, but  with
nontrivial
$R(t), D(t), \gamma (t)  $ obeying certain constraints. Similarly  IHNLS proposed in
\c{PRL05} can be  seen   to be  derivable from  (\ref{ihnls3})
as a particular case   with $D=1, R(t)=g_0e^{c t},
s=1, a=0,  \gamma =0, $ giving $\alpha=0 $, but $\Omega =-c^2 $.
 Therefore both these inhomogeneous NLS \c{PRL00,PRL05} are
equivalent to the standard NLS and hence completely integrable.

  Some  integrable  discrete versions of IHNLS,
 namely inhomogeneous Ablowitz-Ladik models
(ALM)  were proposed  in \c{PRA91,Kon93}, 
  containing  in  addition to  the standard ALM \c{ALM} an inhomogeneous term $ n \omega \psi_n$, with 
$\omega =1 $ \c{PRA91} or $\omega (t) $ as an arbitrary function \c{Kon93} .
 We find that in spite of
 the discrete case a similar reasoning found here holds true  and the
proposed inhomogeneous ALM  can be shown to be  
 gauge equivalent to the standard ALM
\c{ALM}, under discrete
gauge transformation: $ \tilde U_n=g_{n+1}U_ng_n^{-1}, \ \tilde
V_n=g_{n}V_ng_n^{-1}+\dot g_n g_n^{-1},$ with $ g_n=e^{-in\Gamma (t) \sigma^3}$ and redefinition
of the field as $q_n \to \psi_n=q_n e^{i(2n+1)}\Gamma (t) $, where $ \Gamma_t
(t)=\omega (t)$ is an arbitrary function as found in \c{Kon93}.

Based on the above result we therefore  conclude that the  general  inhomogeneous
NLS, if integrable, 
 should be of the form (\ref{ihnls3}).  Other forms of integrable IHNLS
are only   its
 particular cases. However all these  
   inhomogeneous NLS  
 are neither new nor  independent integrable systems, but are 
 equivalent to the standard homogeneous NLS, from which all their
integrable structures like Lax pair, N-soliton solutions, infinite number of   
 commuting conserved
quantities etc. can be  obtained  easily through simple mapping. The  time-dependent   soliton
amplitude, shape and velocity as well as the nonisospectral flow
in these inhomogeneous NLS are just an
artifact of the time-dependent 
 coordinate, gauge and  field transformations, needed to get these systems
from the standard NLS.  Therefore
before proposing  any {\it new } integrable inhomogeneous NLS the authors 
should check
whether it can be linked in any way to the general integrable  IHNLS   (\ref{ihnls3}), whose equivalence with
the well known NLS we have proved here.



\begin{thebibliography}{99}
 \bibitem{CLU76} H. H. Chen and C. S.  Liu, Phys. Rev. Lett. 
{\bf 37}, 693 (1976)  
 \bibitem{RB87} R. Balakrishnan,  Phys. Rev.  
{\bf A 32}, 1144 (1985)  
\bibitem{PRA91} R. Scharf and A. R. Bishop, Phys. Rev.  
{\bf A 43}, 6535 (1991);  
\bibitem{Kon93} V. V. Konotop,  Phys. Rev.  
{\bf E 47}, 1423 (1993);
   
 V. V. Konotop, O. A. Chubykalo and L. Vazquez,  Phys. Rev.  
{\bf E 48}, 563 (1993)

 \bibitem{PRL00} V. N. Serkin and  A. Hasegawa, Phys. Rev. Lett. 
{\bf 85}, 4502 (2000)
 \bibitem{PRL05} Z. X. Liang, Z. D. Zhang and W. M. Liu, Phys. Rev. Lett. 
{\bf 94}, 050402 (2005)
 \bibitem{PRL07} V. N. Serkin, A. Hasegawa
 and T. L. Belyaeva,  Phys. Rev. Lett. 
{\bf 98}, 074102 (2007)  
 \bibitem{arxiv08} H. G. Luo et al
, arXiv: 0808.3437
[nlin.PS];

 H. G. Luo et al,
 arXiv: 0807.1192 
[nlin.PS]
 \bibitem{nls}  M. Ablowitz et al,
 Stud. Appl. Math. {\bf 53} 294 (1974) 
 

M. Ablowitz and H. Segur, {\it Solitons and Inverse Scattering Transforms} (SIAM,
Philadelphia, 1981)

 S. Novikov et al , {\it Theory of Solitons} (Consultants Bureau,
NY, 1984)

 \bibitem{ALM}  M. Ablowitz , Stud.  Appl. Math. {\bf 58} 17 (1978) 

 \bibitem{PLA87} N. Joshi,  Phys. Lett  
{\bf A 125}, 456 (1987)  
 
\end{thebibliography}
  \end{document}